\documentclass[12pt,a4paper]{article}
\usepackage{graphicx,psfrag,bbm,latexsym,color,dcolumn,bm,dsfont,bbm,color,mathrsfs,bbold,latexsym,amsfonts,amssymb}
\usepackage{psfrag}
\usepackage{amsmath}
\usepackage{amssymb}
\usepackage{amsfonts}

\topmargin= - 1.0cm

\newcommand{\mb}[1]{{\mathbf #1}}

\title{On Clausius' approach to entropy and analogies in non-equilibrium}

\author{Giovanni Jona-Lasinio\\
Dipartimento di Fisica, Universit\`a di Roma ``La Sapienza'', \\
and Istituto Nazionale di Fisica Nucleare,\\
Piazzale A. Moro 2, Roma 00185, Italy\\
E-mail: gianni.jona@roma1.infn.it}

\date{}
\begin{document}
\maketitle

\begin{abstract}
In his ninth memoir Clausius summarizes the two principles of thermodynamics as follows 
\begin{quotation}
  {\sl The whole mechanical theory of heat rests on two fundamental theorems, - that of equivalence of heat and work, and that of equivalence of transformations.}\end{quotation}
This paper contains an introduction to Clausius' approach to entropy as illustrated in his original articles and describes an analogy in the macroscopic fluctuation theory of non--equilibrium diffusive systems.

\end{abstract}
\section{Introduction}
I often asked myself: {\sl ``How did Clausius get the idea of entropy in a pure thermodynamic approach?'' }.  Entropy is a notion which carries some mystery with it, 
as was emphasized by a qualified opinion. There is a quantity called entropy in information theory which measures  the content of information of a message. A story tells that such a terminology was suggested to Shannon, the inventor of the theory, by the famous mathematician John Von Neumann with the comment 

\begin{quotation}  
{\sl ``You should call it entropy, for two reasons. In the first place your uncertainty function has been used in statistical mechanics under that name, so it already has a name. In the second place, and more important, no one really knows what entropy  is, so in a debate you will always have the advantage.''}
\end{quotation}

\medskip

The fourth, sixth and ninth memoirs of Clausius by the titles {\sl ``On a modified form of the second fundamental theorem of the mechanical theory of heat'' \cite{c4}; ``On the application of the theorem of the equivalence of transformations to interior work'' \cite{c6}; ``On several convenient forms of the fundamental equations of the mechanical theory of heat'' \cite{c9}} respectively, are very illuminating, however for some reason his arguments did not attract so much attention and are not reported in the treatises on thermodynamics I know.

\medskip
 
Clausius idea, as developed in his fourth memoir, is that there must exist a quantity called the {\sl the equivalence value of a transformation} expressing the strict connection of a transformation of heat into work with transfer of heat from a higher temperature to a lower one as in a Carnot cycle. The equivalence value is  what he will call later the entropy. All the reversible transformations connecting two given states have the same equivalence value, so that entropy is an invariant under variation of the protocols connecting those states.

\medskip

The existence of  entropy in classical equilibrium thermodynamics implies that the integral $\int {\frac {\delta Q}{T}}$ does not depend on the particular reversible quasi-static transformation joining two fixed initial and final states. For example in the simple case of an ideal gas we have the exact 1-form
\begin{equation}
dS = C_p {\frac {dT}{T}} - nR{\frac {dp}{p}}
\end{equation}  

\medskip

In this note, after illustrating Clausius' reasoning \footnote {I became aware only recently of the paper \cite{xue} where Clausius approach is also reviewed.}, we wish to show that a formula obtained in \cite{mft,qc} represents a generalization to stationary non-equilibrium states in the context of an important class of irreversible processes. These are the diffusive processes considered in the macroscopic fluctuation theory (MFT).

\medskip

The  macroscopic fluctuation theory deals mainly with diffusive systems boundary driven and in presence of external fields. For these systems a finite time approach to macroscopic transformations has been developed \cite{mft,qc}. Quasi static transformations are obtained as limits of scaled hydrodynamic equations with suitable time dependent boundary conditions.

\medskip

A quasi-static transformation is a continuous succession of stationary states modified by  $1/\tau$ corrections, where $\tau$ is a large time. Then one takes  the limit $\tau \rightarrow \infty$. Furthermore new exact expressions connecting the equilibrium free energy or its derivatives to deviations from stationarity follow from a systematic development in powers of $1/\tau$.

\medskip

The fourth memoir of Clausius \cite{c4} clarifies how the idea of entropy emerged thermodynamically. He concentrates on transformations rather than states and wants to characterize the relationship between transforming heat into work and the transition of heat between two temperatures in a generalized Carnot's cycle. The leading idea is that there exists a quantity called {\sl the equivalence value of a transformation}, characterising these different changes of state. He then gives a general argument to show that such a quantity must have the form $\int dQ/T$. 

\medskip

Let us quote from his fourth memoir

\medskip

\begin{quotation}
{\sl ``Carnot's theorem,....,expresses a relation between two kinds of transformations, the transformation of heat into work, and the passage of heat from a warmer to a colder body, which may be regarded as the transformation of heat at the higher, into heat at a lower temperature.....

For example, let the quantity of heat $Q$, produced in any manner whatever from work, be received by the body $K$; then by the foregoing cyclical process [he considers a generalized Carnot's cycle] it can be withdrawn from $K$ and transformed back into work, but at the same time the quantity of heat $Q_1$ will pass from $K_1$ to $K_2$; or if the quantity of heat $Q_1$ had previously been transferred from $K_1$ to $K_2$, this can be again restored to $K_1$ by the reversed cyclical process wherby the transformation of work into the quantity of heat $Q$ of the temperature of the body $K$ will take place.

We see, therefore, that these two transformations may be regarded as phenomena of the same nature, and we may call two transformations which can thus mutually replace one another {\emph equivalent}. We have now to find the law according to which the transformations must be expressed as mathematical magnitudes, in order that the equivalence of two tranformations may be evident from the equality of their values. The mathematical value of a transformation thus determided may be called its {\emph equivalence-value} (Aequivalenzwerth).'' }  
\end{quotation}

\medskip

Here is  Clausius' argument to determine such equivalence-value

\medskip

\begin{quotation}
{\sl ``With respect to the magnitude of the equivalence-value, it is first of all clear that the value of a transformation from work into heat must be proportional to the quantity of heat produced; and besides this it can only depend upon the temperature. Hence the equivalence-value of the transformation of work into the quantity of heat $Q$, of the temperature $t$, may be represented generally by
\begin{equation}  
Q\, f(t)
\end{equation}
wherein $f(t)$ is a function of the temperature which is the same for all cases....In a similar manner the value of the passage of the quantity of heat $Q$, from the temperature $t_1$ to the temperature $t_2$, must be proportional to the quantity $Q$, and besides this, can only depend on the two temperatures. In general therefore, it may be expressed by
\begin{equation}
Q\, F(t_1,t_2)
\end{equation}
wherein $F(t_1,t_2)$ is a function of both temperatures, which is the same for all cases, and of which we at present only know that, without changing its numerical value, it must change its sign when the two temperatures are interchanged; so that
\begin{equation}
F(t_1,t_2)=-F(t_2,t_1)
\end{equation}''}
\end{quotation}

\medskip

Clausius concludes that the relationship between $f(t)$ and $F(t_1,t_2)$ is
\begin{equation}
F(t_1,t_2)=f(t_2)-f(t_1)
\end{equation}
He writes $f(t)=\frac {1}{T}$ where $T$ is provisionally a function of temperature that will be identified later with the absolute temperature. In his ninth memoir \cite{c9} he will summarize the two principles of thermodynamics as follows

\medskip

\begin{quotation}
{\sl ``The whole mechanical theory of heat rests on two fundamental theorems, - that of equivalence of heat and work, and that of equivalence of transformations.''}\end{quotation}

\medskip

Some comments are here in order. The equivalence of transformations in the above reasoning is restricted to heat going into work and heat passing from higher temperature to lower temperature. This is equivalent to the second principle. However once the existence of entropy is established it follows that to all reversible transformations between given initial and final states of a system can be attributed the same equivalence value i.e. the difference of entropies between initial and final state.

\medskip

We now show that an analogy exists in the dissipative context of non--equilibrium diffusive systems. To make the paper selfcontained we collect here some facts and formalism of the macroscopic fluctuation theory, a kind of ritual, which can be found in greater detail in \cite{mft,qc} and references therein. 

\medskip

\section{\sl Macroscopic Dynamics}
 We denote by $\Lambda \subset \mb R^d$ the
bounded region occupied by the system, by $\partial \Lambda$ the
boundary of $\Lambda$, by $x$ the macroscopic space coordinates and by
$t$ the macroscopic time. The system is in contact with boundary
reservoirs, characterized by their chemical potential $\lambda (t,x)$, $x \in \partial \Lambda$ and under the action of an external field $E(t,x)$.

\medskip

At the macroscopic level the assumption is that the system is completely described by the local density $\rho(t,x)$ and the local density current $j(t,x)$ and their evolution is given by the continuity equation together with the constitutive equation which expresses the current as a function of the density. Namely,
\begin{equation}
\label{2.1}
\begin{cases}
\partial_t \rho (t) + \nabla\cdot j (t) = 0,\\
j (t)= J(t,\rho(t)),
\end{cases}
\end{equation}
where we omit the explicit dependence on the space variable $x\in\Lambda$.
For driven diffusive systems the constitutive equation takes the form
\begin{equation}
\label{2.2}
J(t,\rho)  = - D(\rho) \nabla\rho + \chi(\rho) \, E(t),
\end{equation}
where the \emph{diffusion coefficient} $D(\rho)$ and the \emph{mobility}
$\chi(\rho)$ are assumed to be $d\times d$ symmetric and positive definite matrices. This holds in the context of stochastic lattice gases \cite{mft}. Equation \eqref{2.2} relies  on the local equilibrium hypothesis, small local gradients and linear response to the external field. The evolution of the density is thus given  by the driven diffusive equation
\begin{equation}
\label{r01}
\partial_t \rho (t) + \nabla\cdot \big( \chi(\rho)  E(t) \big)
= \nabla\cdot \big( D(\rho) \nabla\rho \big).
\end{equation}
The transport coefficients $D$ and $\chi$ satisfy the local Einstein relation
\begin{equation}
\label{r29}
D(\rho) = \chi(\rho) \, f''(\rho),
\end{equation}
where $f$ is the equilibrium free energy per unit  volume which we assume
to depend on the local value  $\rho (x)$.

\medskip

Equations \eqref{2.1}--\eqref{2.2} have to be supplemented by the
appropriate boundary conditions on $\partial\Lambda$ due to the
interaction with the external reservoirs. If $\lambda(t,x)$,
$x\in\partial \Lambda$, is the chemical potential of the external
reservoirs, the boundary condition reads
\begin{equation}
\label{2.3}
f'\big(\rho(t,x) \big) = \lambda(t,x), \qquad x\in\partial \Lambda.
\end{equation}

\medskip

If the chemical potential and external field do not depend on time, we
denote by $\bar\rho=\bar\rho_{\lambda, E}$ the stationary solution of
\eqref{r01},\eqref{2.3},
\begin{equation}
\label{05}
\begin{cases}
\nabla \cdot J(\bar\rho)= \nabla \cdot \big( -D(\bar\rho)
\nabla\bar\rho + \chi(\bar\rho) \, E  \big) = 0,  
\\
 f' (\bar\rho(x)) = \lambda (x), 
\qquad x\in\partial \Lambda.
\end{cases}
\end{equation}
We will assume that this stationary solution is unique. The stationary density profile $\bar\rho$ is characterized by the vanishing of the divergence of the associated current, $\nabla \cdot J(\bar\rho)=0$.  A special situation is when the current itself vanishes, $J(\bar\rho)=0$; if this is the case, which means that the boundary conditions and the external field balance each other, we say that the system is in a macroscopic equilibrium state; this can be viewed as a macroscopic counterpart to detailed balance \cite{tt}. 

\medskip

Homogeneous equilibrium states correspond to the case in which the external field vanishes and the chemical potential is constant in space.
Inhomogeneous equilibrium states correspond to the case in which the external field is gradient, $E=-\nabla U$, and it is possible to choose the arbitrary constant in the definition of $U$ such that $U(x)=-\lambda(x)$, $x\in\partial\Lambda$. The stationary equation $\nabla \cdot J = 0$ with the associated boundary conditions play a role akin to the equation of state in equilibrium thermodynamics.

\medskip

\section{\sl Transformations and Energy Balance}

Consider a system in a time dependent environment, that is, $E$ and $\lambda$ depend on time. The work done by the environment on the system in the time interval $[0,T]$ is
\begin{equation}
  \label{W=}
  \begin{split}
    W_{[0,T]} = & \int_{0}^{T}\! dt \, \Big\{ \int_\Lambda \!dx\, j(t)
    \cdot E(t)
    - \int_{\partial\Lambda} \!d\sigma \, \lambda
    (t) \: j(t) \cdot \hat{n} \Big\},
  \end{split}
\end{equation}
where $\hat n$ is the outer normal to $\partial \Lambda$ and $d\sigma$
is the surface measure on $\partial \Lambda$.  The first term on the
right hand side is the energy provided by the external field
while the second is the energy provided by the reservoirs.

\medskip

Fix time dependent paths $\lambda(t)$ of the chemical potential and
$E(t)$ of the field. Given a density profile $\rho_0$, let
$\rho(t)$, $j(t)$, $t \ge 0$, be the solution of
\eqref{2.1}--\eqref{2.3} with initial condition $\rho_0$.  
By using the Einstein relation \eqref{r29} and the boundary condition 
$f'(\rho(t)) = \lambda(t)$, an application of the
divergence theorem yields
\begin{equation}
\label{04}
W_{[0,T]} \,=\,  F(\rho(T)) - F(\rho(0)) 
+\,  \int_{0}^{T} \!dt  \int_\Lambda \!dx \;
j(t)\cdot \chi(\rho(t) )^{-1} j(t),
\end{equation}
where $F$ is the equilibrium free energy functional,
\begin{equation}
\label{10}
F(\rho) = \int_\Lambda \!dx \: f (\rho(x)).
\end{equation}
The step from \eqref{W=} to \eqref{04} uses  the constitutive equation  \eqref{2.2} and the Einstein relationship \eqref{r29}. It tells us that for the class of systems considered the change of the free energy $\Delta F$ is, as expected, the difference between the total work and the total dissipation. Notice that if the system is out of equilibrium for an infinite time both these quantities are infinite. The currents $j(t)$ must be evaluated on the solutions of the hydrodynamic equations.

\medskip

From \eqref{04} follows the Clausius inequality
\begin{equation}
W_{[0,T]} \,\geq\,  F(\rho(T)) - F(\rho(0)) 
\end{equation}
For quasi-static transformations between nonequilibrium stationary states, the Clausius inequality  does not carry any significant information. In fact, the energy dissipated along such transformations will necessarily include the contribution needed to maintain the nonequilibrium stationary states, which is infinite in an unbounded time window. It is however possible to formulate a meaningful version of the inequality for nonequilibrium states by introducing a \emph{renormalized work} $W^\mathrm{ren}$ that is defined by subtracting the energy needed to maintain the nonequilibrium state \cite{mft}. This energy interprets in the present context what is called {\sl housekeping heat} in \cite{opa}

\medskip

To analyze transformations over a long interval $[0,\tau]$ driven by slowly changing boundary conditions and external fields, it is convenient to rescale time and introduce the dimensionless variable $s={t}/{\tau}$. The \emph{protocol} of a transformation  is defined therefore by a choice of the external drivings $E(s,x)$, $x\in\Lambda$, and $\lambda(s,x)$, $x\in \partial\Lambda$, $s\in[0,1]$.

\medskip

Introduce the expansion
\begin{equation}\label{davide}
\begin{array}{l}
\displaystyle{
\rho^\tau (\tau s)=\bar \rho(s)+\tfrac 1\tau \, r(s)
+ o\big(\tfrac1{\tau}\big)\,,
} \\
\displaystyle{
j^\tau(\tau s)
=J(s,\bar\rho(s))+\tfrac 1\tau \, g(s) + o\big(\tfrac 1{\tau} \big)
\,.}
\end{array}
\end{equation}
It is easy to see that $r,g$ obey the following equations
\begin{equation}\label{davide2bis}
\left\{
\begin{array}{l}
\partial_s  \bar \rho(s) + \nabla \cdot g(s) =0 \\
g(s)= -
\nabla\cdot\big( D(\bar\rho(s)) r(s) \big)
+ r(s) \chi'(\bar \rho(s)) E(s)
\\
r(s, x)=0, \; x\in\partial \Lambda
\end{array}
\right.
\end{equation}
which implies that $r,g$ depend linearly on $\partial_s \bar\rho$.

\medskip

The formula we want to discuss is
\begin{equation}
\label{f}
  \begin{split}
F\big(\bar \rho(1)\big)-F\big(\bar \rho(0)\big)
    &\; = \int_0^1\!ds \int_\Lambda\!dx \, E(s)\cdot g(s) -\int_0^1\!ds
    \int_{\partial \Lambda}\! d\sigma \,\lambda(s) g(s)\cdot \hat n
    \\
    &\;\; + \int_0^1\!ds \int_\Lambda \!dx\, r(s)
    J(s,\bar\rho(s))\cdot (\chi^{-1})'\big(\bar \rho(s)\big)
    J(s,\bar\rho(s))
    \,.
  \end{split}
\end{equation}
For a derivation we refer to \cite{mft,qc}.

\medskip

Taking the pair $E (x,s),\lambda (x,s)$ as our state variables, a transformation corresponds to a path $\gamma$ in an infinite dimensional space. Then the right hand side of \eqref{f} can be represented as a line integral of an exact 1-- form
\begin{equation}
\int_{\gamma} A_1 \, dE + A_2 \, d\lambda \,.   
\end{equation}
We now take into account that $\bar\rho$ depends on time only through $\lambda,E$ 
\begin{equation} 
\partial_s \bar\rho = {\frac {\delta \bar\rho}{\delta \lambda}} \partial_s \lambda + {\frac {\delta \bar\rho}{\delta E}} \partial_s E \,.
\end{equation}
Inserting this expression into the formulas for $r,g$  in \eqref{f}, separating $\partial_s \lambda$ and $\partial_s E$ we can obtain explicitly $A_1, A_2$.

\medskip

Equation \eqref{f} establishes a mathematical equivalence of quasi--static transformations for a class of dissipative systems. These transformations can be energetically  very different as the energy necessary to keep the system out of equilibrium per unit of time can differ considerably. Concrete examples will be discussed elsewhere.

\section{Concluding remarks}
For diffusive systems  different equivalent transformations in the sense of \eqref{f} can be very different as far as energy consumption is concerned. To be concrete let us consider a conductor in a potential difference $\Delta V$ that we want to increase to a value $\Delta V'$. A quasi--static transformation increasing slowly
$\Delta V$ would produce a very large quantity of heat, infinite in the limit, that we can possibly reduce through the following procedure that we formulate in a slightly more general form. 

\medskip

Consider a quasi-static transformation between the stationary states corresponding to the values $E_i , \lambda_i$ and $E_f , \lambda_f$. You may ideally split the transformation in the following way. 

\medskip

Change abruptly the field from $E_i$ to $E_i^{'}$ where $E_i^{'}$ is the value corresponding to equilibrium, i.e. $J=0$, with chemical potential $\lambda_i$. The system will relax to equilibrium. Then perform a quasi-static transformation via equilibrium states until you reach the final value $\lambda_f$. No housekeeping heat. Finally bring  abruptly the field to its final value $E_f$. The system will relax to its final stationary state. 

\medskip

Of course in every concrete case one must estimate the energy cost of the splitting and decide whether over a long but finite interval of time of a real transformation it is convenient. In \cite{qc} we discussed a criterion of optimality based on the minimisation of the renormalized work which led to some unexpected outcome also for a system of free particles.

\medskip

In conclusion, it is somewhat surprising that an equivalence class of transformations could be defined for irreversible dissipative systems like purely diffusive systems. It is interesting to investigate the possibility of generalising this analysis to a more general setting like reaction--diffusion macroscopic dynamics where presumably we can extend the macroscopic fluctuation theory.

\subsection*{Acknowledgements}
I wish to acknowledge my long standing collaboration with L. Bertini, A. De Sole, D. Gabrielli, C. Landim which ispired this paper.


\begin{thebibliography}{99}

\bibitem{c4}
Clausius R.,  
\emph{\sl On a modified form of the second fundamental theorem of the mechanical theory of heat}, Pogg. Annalen, {\bf xciii} 481 (1854); translated in Phil. Mag.  {\bf xii} 81 (1856).

\bibitem{c6}
Clausius R. , \emph{\sl ``On the application of the theorem of the equivalence of transformations to interior work''}, Pogg. Annalen, {\bf cxvi} 73 (1862); translated in Phil. Mag. S.4. {\bf xxiv} 81 (1862). 

\bibitem{c9}
Clausius R.;
\emph{On several convenient forms of the fundamental equations of the mechanical theory of heat}, Pogg. Annalen {\bf cxxv}  353 (1865); translated in Journ. de Liouville {\bf 10} 361 (1865).  
 
\bibitem{mft}
 Bertini L., De Sole A., Gabrielli D., Jona-Lasinio G., Landim C.; 
  \emph{Macroscopic fluctuation theory}, Rev. Mod. Phys. {\bf 87} 593 (2015), ArXiv:1404.6466.

\bibitem{qc}
Bertini L., De Sole A., Gabrielli D., Jona-Lasinio G., Landim C.; 
\emph{Quantitative analysis of Clausius inequality}, J. Stat. Mech. Theory Exp. P10018 (2015).

\bibitem{tt}
 Bertini L., De Sole A., Gabrielli D., Jona-Lasinio G., Landim C.; \emph{Towards a non--equilibrium thermodynamics: a self--contained macroscopic description of driven diffusive systems}, J. Stat. Phys. {\bf 135}, 857 (2009).
   
\bibitem{opa}
Oono Y., Paniconi M., \emph{Steady state thermodynamics}, Progr. Theor. Phys. Suppl. {\bf 130}, 29 (1998).   

\bibitem{xue}
Ti-Wei Xue, Zeng-Yuan Guo, \emph{What is the Real Clausius Statement of the Second Law of Thermodynamics?}, Entropy {\bf 2019}, 21, 926.
  


\end{thebibliography}
\end{document}